\documentclass[lettersize,journal]{IEEEtran}
\usepackage{amsmath,amsfonts}
\usepackage{algorithmic}
\usepackage{array}
\usepackage[caption=false,font=normalsize,labelfont=sf,textfont=sf]{subfig}
\usepackage{textcomp}
\usepackage{stfloats}
\usepackage{url}
\usepackage{verbatim}
\usepackage{graphicx}
\usepackage{hyperref}
\graphicspath{{figs/}{figures/}{pictures/}{images/}{./}} %

\usepackage{tabu}                      %
\usepackage{booktabs}                  %
\usepackage{lipsum}                    %
\usepackage{mwe}                       %

\usepackage{diagbox}

\usepackage{mathptmx}                  %
\usepackage{amsmath,amsfonts}
\usepackage{algorithmic}
\usepackage{algorithm}
\usepackage{array}
\usepackage{textcomp}
\usepackage{stfloats}
\usepackage{url}
\usepackage{verbatim}
\usepackage{graphicx}
\usepackage{cite}  
\usepackage{amsmath}
\usepackage{xcolor}

\usepackage{tabularx}  %
\usepackage{multirow}  %
\usepackage{booktabs}  %

\def\BibTeX{{\rm B\kern-.05em{\sc i\kern-.025em b}\kern-.08em
    T\kern-.1667em\lower.7ex\hbox{E}\kern-.125emX}}
\usepackage{balance}
\begin{document}
\title{\emph{Contextualization or Rationalization?} The Effect of Causal Priors on Data Visualization Interpretation}
\author{Arran Zeyu Wang, David Borland, Estella Calcaterra, and David Gotz
  \thanks{A. Z. Wang, E. Calcaterra, and D. Gotz are with the University of North Carolina at Chapel Hill.} 
  \thanks{D. Borland is with RENCI and the University of North Carolina at Chapel Hill.}}

\markboth{Journal of \LaTeX\ Class Files,~Vol.~18, No.~9, September~2020}%
{How to Use the IEEEtran \LaTeX \ Templates}

\renewcommand{\sectionautorefname}{Section}
\renewcommand{\subsectionautorefname}{Section}
\renewcommand{\subsubsectionautorefname}{Section}

\newcommand{\fix}[1]{\textcolor{red}{{{#1}}}}
\newcommand{\add}[1]{\textcolor{black}{#1}}
\newcommand{\new}[1]{\textcolor{black}{#1}}

\maketitle

\begin{abstract}
Understanding how individuals interpret charts is a crucial concern for visual data communication. This imperative has motivated a number of studies, including past work demonstrating that causal priors---\emph{a priori} beliefs about causal relationships between concepts---can have significant influences on the perceived \emph{strength of variable relationships} inferred from visualizations.  This paper builds on these previous results, demonstrating that causal priors can also influence the \emph{types of patterns} that people perceive as the most salient within \textit{ambiguous scatterplots} that have roughly equal evidence for trend and cluster patterns.
\add{Using a mixed-design approach that combines a large-scale online experiment for breadth of findings with an in-person think-aloud study for analytical depth, we investigated how users' interpretations are influenced by the interplay between causal priors and the visualized data patterns.}
\add{Our analysis suggests two archetypal reasoning behaviors through which people often make their observations: \emph{contextualization}, in which users accept a visual pattern that aligns with causal priors and use their existing knowledge to enrich interpretation, and \emph{rationalization}, in which users encounter a pattern that conflicts with causal priors and attempt to explain away the discrepancy by invoking external factors, such as positing confounding variables or data selection bias.}
\add{These findings provide initial evidence highlighting the critical role of causal priors in shaping high-level visualization comprehension, and introduce a 
\textit{vocabulary} for describing how users reason about data that either confirms or challenges prior beliefs of causality.}
\end{abstract}


\begin{figure*}
    \centering
\includegraphics[width=\textwidth]{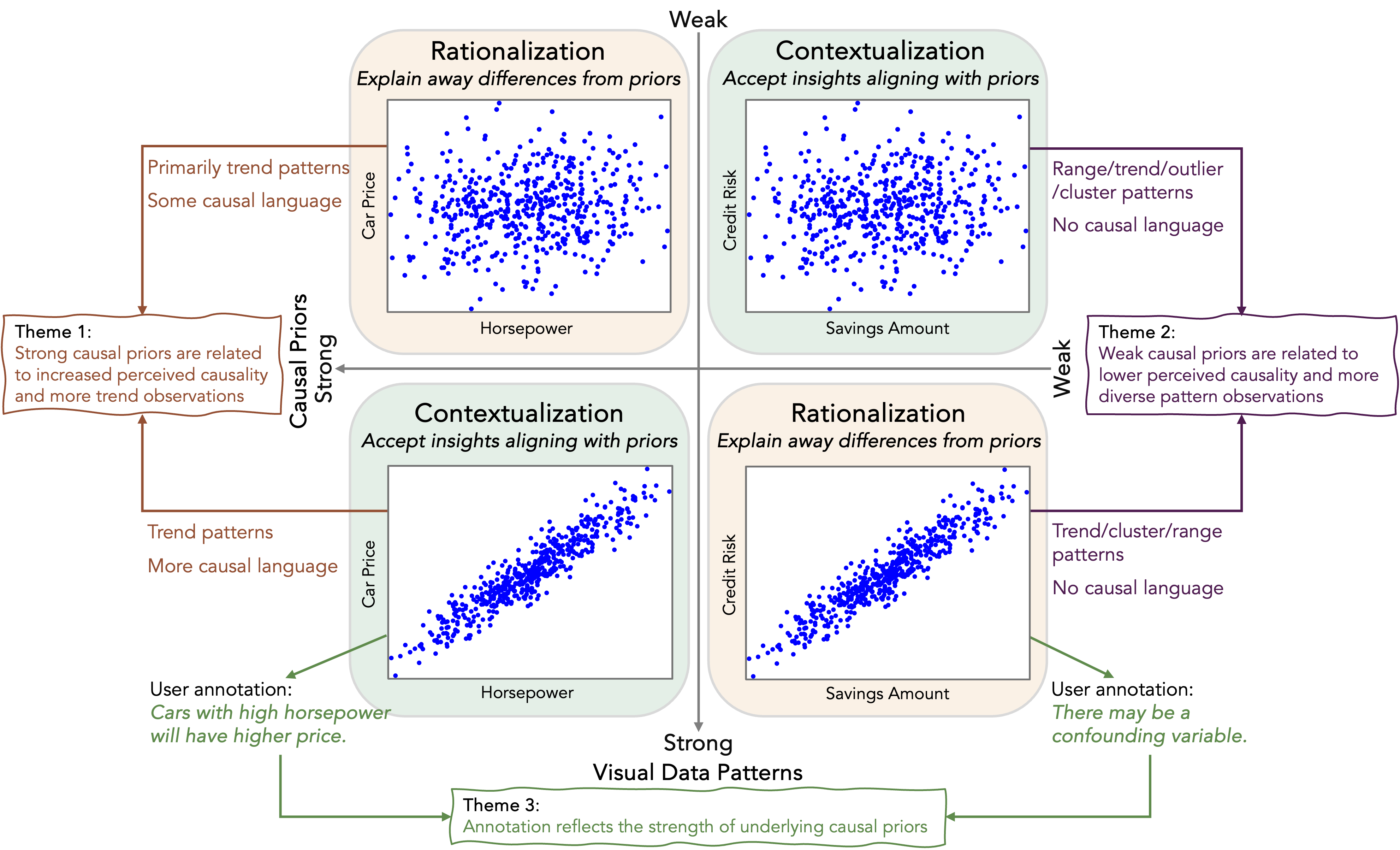}
  \vspace{-1em}
\caption{ 
\add{A schematic representation of the primary insights from our experiments. The X-axis represents the strength of \textit{causal priors} with stronger priors on the left. The Y-axis represents the strength of \textit{visual patterns} with stronger patterns on the bottom.
An illustrative scatterplot stimulus is shown in each quadrant.
The figure highlights key insights from our study, including (1) 
two observed archetypal data visualization interpretation behaviors---\emph{contextualization} and \emph{rationalization}---driven by the interplay between causal prior and visualization pattern strength; and (2) the influence of causal priors on visualization interpretations as described in Themes 1, 2, and 3.
}}
  \vspace{-1em}
\label{fig:teaser}
\end{figure*}

\section{Introduction}
\label{sec-intro}

Data visualization has become an indispensable tool for data exploration and analysis, yet our understanding of the interpretation and comprehension of visual representations often lacks theoretically and empirically grounded foundations~\cite{munzner2014visualization, franconeri2021science, szafir2023visualization}.
Much prior empirical research has focused on graphical perception~\cite{cleveland1984graphical}, examining how low-level graphical features reflecting patterns in data are perceived by users. Although necessary to understand how individuals process visualizations, such experiments do not consider much of the contextual information, including prior beliefs and expectations, that are involved in visualization comprehension~\cite{evans2013rationality, nickerson1998confirmation}.
This broader context through which a user interprets a visualization can be crucial, as it can fundamentally affect how people interpret the graphical representations they encounter. 

Studies that have examined higher-level chart interpretation and comprehension---encompassing reasoning, sensemaking, and mental processes---have shown that the ways in which people comprehend charts are complicated and influenced by many factors.
For instance, a recent study demonstrated that different people viewing the same chart can arrive at widely different interpretations~\cite{bearfield2024same}. Other work has also shown that 
users' interpretations can be unpredictable and often do not align with a designer’s objectives~\cite{quadri2024do}, and that interpretations vary depending on an individual's mental model~\cite{williams2023data}. 
The unpredictability of the insights that people draw from the relatively simple charts used in these studies suggests that our understanding of the many factors that influence users as they make observations from visualized data patterns remains far from complete.

One underexplored factor that can influence chart interpretation is the presence of \emph{causal priors}: pre-existing beliefs about the causal relationships between concepts depicted in a visualization. Previous work has shown that differences in causal priors can significantly influence users' judgments of causality from visualized data~\cite{wang2024causal}.
The results provide compelling evidence that prior assumptions about causal relationships between the variables shown in a chart comprise part of the broader context through which the chart is understood.  However, that study focused solely on how causal priors influence the strength of causal relationships inferred from visualizations. Previous work has not yet examined how causal priors might influence other factors related to visualization comprehension.

This paper \add{describes initial work to} help bridge this gap by focusing on the impact of causal priors on  high-level interpretation of visualizations
We focus on scatterplots within which we simultaneously display two common visual data patterns---\textit{trends} and \textit{clusters}.
Moreover, we focus on ambiguous charts that have a relatively balanced perceived strength for each pattern. This is because charts with a single dominant visual pattern are likely to be interpreted in a similar way by all users~\cite{quadri2024do}. In contrast, 
ambiguous visual representations allow for different visual data patterns that could be plausibly identified by users as the main pattern in the chart. Focusing on ambiguous charts in our study design enables us to explore the effect of causal priors by examining which patterns are identified by users as the dominant visual data pattern in a visualization.
\add{To investigate this question, we conducted a mixed-design study} consisting of both crowdsourced and in-person experiments. This design allowed us to capture both broad data at scale as well as more detailed evidence from in-person participants (\autoref{fig:exp}).

\add{In the crowdsourced experiment, we collected a broad base of qualitative feedback to identify the general effects of causal priors on visualization interpretation.
In the in-person experiment, we explored more deeply how people make sense of these visualizations via a think-aloud protocol and by asking them to make annotations to help others understand their observations.}
We collected and qualitatively analyzed users' reported \emph{observations}, \emph{informative visual data patterns}, \emph{annotations}, and \emph{reasoning strategies} from visualizations (\autoref{sec:analysis}).
Interpretations varied for different causal priors, even when shown with the same visualized data. The three main themes uncovered by our analysis (shown in \autoref{fig:teaser}) are: i) \textcolor[RGB]{148, 80, 52}{strong causal priors are associated with increased perceived causality and observations of trends} (\autoref{sec:t1}); ii) \textcolor[RGB]{79, 28, 81}{weak causal priors are associated with lower perceived causality and more diverse pattern observations} (\autoref{sec:t2}); and iii) \textcolor[RGB]{95, 139, 76}{annotations reflect the strength of underlying causal priors} (\autoref{sec:t3}).
\new{Our results suggest that the perceived causal relationship between variables is modulated by the strength of the user's pre-existing causal prior, and the visual data pattern a user identifies as primary is influenced by the alignment between pattern strength and causal prior strength.}

Further, from users' reasoning strategies, we characterized two archetypal interpretation behaviors: \textit{contextualization} and \textit{rationalization}. \new{Contextualization occurs when users readily integrate visual patterns that align with their expectations into their existing causal understanding, whereas rationalization occurs when users discount or explain away patterns that contradict their beliefs, often by invoking external factors} (see coordinate quadrants in \autoref{fig:teaser}).
Contextualization usually appears when the variables' causal priors and visualization's visual data patterns are congruent, i.e., both strong or both weak. With this behavior, \emph{the information shown in a visualization may help individuals ground their interpretation within the context of their causal priors} (\autoref{sec:b1}).
Essentially, given ambiguous visual cues, they focus on those that align with their prior beliefs.
For example, if someone already believes that two variables are causally linked, they might overlook the presence of clusters in the visualization, instead focusing on evidence for a trend in the data that supports the causal prior.

On the other hand, rationalization is more common when causal priors and visual data patterns are incongruent, i.e., have different relative strengths. With this behavior, \emph{users propose reasons, usually external, to attempt to explain the differences between the visualization and their causal priors} (\autoref{sec:b2}).
They rationalize the ambiguous cues in visualizations that differ from their expectations by imagining additional factors, such as lack of data, additional confounders and colliders, or other unseen variables. This leads them to discount their observations and articulate increased uncertainty.
Rather than settling on a clear narrative, individuals speculate on various possible underlying causes to explain the visual data patterns, commonly relying on assumptions beyond the scope of the chart, reflecting diverse observations and visualization interpretations in line with causal priors.

Our findings contribute to further understanding of the role of causal priors in people's interpretation of data visualizations. Specific contributions include:
\begin{itemize}
    \item \textbf{A mixed design study including both crowd-sourced and in-person participants}. Our design includes crowd-sourced participants for more general insights and in-person participants for deeper explorations.
    \item \textbf{Applying the lens of causality on visualization interpretation}. Our study explores the long-standing question of how visualizations are interpreted via a novel lens of causal priors, and demonstrates that preconceived causal priors are associated with bias in the processing of information, resulting in visualized data being selectively and subjectively interpreted to fit existing priors.
    \item \textbf{Characterization of two archetypal visualization interpretation behaviors}. Our results identify and characterize two visualization interpretation behaviors---contextualization and rationalization---that were observed for different combinations of the strengths of prior beliefs and visualized patterns.
\end{itemize}

\section{Background}
\label{sec-related}

Our work builds upon and intersects with several established research topics. Here we review relevant previous work in graphical perception, high-level interpretation and comprehension of data visualizations, and the perception of causality in visualized data.

\subsection{Graphical Perception}

Prior research in data visualization has long focused on exploring how information is processed and interpreted~\cite{szafir2023visualization}.
Early work by Cleveland and McGill established foundational principles for graphical perception~\cite{cleveland1984graphical}, with various researchers conducting a large number of visualization-specific follow-up studies (see Quadri \& Rosen~\cite{quadri2021survey} for a survey). Such studies highlight that visualization task accuracy is influenced by both design and individual interpretation strategies.
Graphical perception experiments typically focus on specific aspects of visualization comprehension, such as finding a piece of statistical information~\cite{sarikaya2017scatterplots} or judging different patterns~\cite{wilkinson2005graph}, and they often focus on specific visual encoding channels or representations, such as color~\cite{szafir2018modeling}, shape~\cite{tseng2024shape}, and composition~\cite{gleicher2011visual}.
Across all these experiments, scatterplots are the most widely employed proxy to explore general insights~\cite{sarikaya2017scatterplots, szafir2023visualization, rensink2013prospects}.

However, the majority of such work focused on low-level tasks~\cite{quadri2021survey}, instead of higher-level comprehension and interpretation.
Research connecting cognitive psychology with data visualization~\cite{elliott2020design, wang2025characterizing, bae2025bridging} has established that the accuracy with which users interpret visualized data is not only a function of the chart design and visual encoding but also of the more complex cognitive processes involved in decoding the information.

\subsection{High-Level Comprehension in Visualization}

Beyond low-level graphical perception tasks, initial efforts have attempted to directly measure and estimate visualization interpretations, which is crucial for data visualization communication~\cite{franconeri2021science}.
In a data abstraction study, the authors noted that people's mental models and data abstractions of the same dataset presentations can initially be diverse~\cite{williams2023data}.
A psychology study~\cite{zacks1999bars} found that individuals focus on different features when asked to describe a graph, which in turn influences the conclusions they draw about the data; however, they only studied the simplest two-point graphs.
Similarly, human subjects' reasoning results on the same question when looking at the same visualization were also found to vary~\cite{bearfield2024same}.
However, these tasks centered around how people judge and reason for specifically designed questions and predictions without exploring their general interpretation and comprehension.
Studies have also reported that consumers of visualizations may have different understandings based on various impact factors, such as the curse of knowledge~\cite{xiong2019curse}, political events~\cite{yang2023swaying}, rhetorical framing~\cite{hullman2011visualization}, users' background~\cite{peck2019data}, \add{public receptivity~\cite{he2023enthusiastic}}, data literacy~\cite{burns2023we}, text~\cite{stokes2022striking}, and even language describing intent~\cite{adar2020communicative}.

Perhaps most closely related to our study is an empirical experiment focused on visualization comprehension~\cite{quadri2024do}. This work challenged some results from previous perception studies and design guidelines, finding that visualization comprehension does not always align with the designers' stated objective. Moreover, it found that results from traditional perception experiments were often not able to effectively predict what knowledge people could learn from a graph.
These limitations contribute to the motivation for this paper's focus on understanding the effects of preexisting causal beliefs on visualization interpretation.

\subsection{The Perception of Causality from Visualized Data}

Efficiently conveying causal inference is another emerging topic in the data visualization community~\cite{borland2024using, hullman2021designing}.
Recently, theoretical and empirical contributions demonstrated that, while visualizations typically do not directly represent causal relationships, people tend to make causal judgments even with very simple visualizations~\cite{xiong2019illusion, kale2021causal, wang2024causal, wang2024empirical, wang2024beyond}.
More specifically, previous work has reported that chart types, such as bar charts~\cite{xiong2019illusion} and scatterplots~\cite{wang2024empirical}, may differ in the extent to which users infer causality, and visual features such as trends~\cite{xiong2019illusion} and associations~\cite{wang2024causal} can lead to users making causal interpretations.
However, users' causal inferences may be relatively insensitive to sample size~\cite{kale2021causal}, and may be heavily impacted by preexisting factors external to the visualization, such as causal priors between variables~\cite{wang2024causal}.
\add{Another thread of related work has explored how users update their beliefs about correlation through a Bayesian cognitive framework~\cite{karduni2020bayesian}, with a focus on persuasion~\cite{markant2023data} and attitude~\cite{rogha2024impact}. These studies demonstrate a mechanism for how judgments are shaped by incoming data.}

\add{Our work complements these findings by focusing not on the process of belief updating itself, but on how initial causal priors influence the types of patterns and features individuals identify as most salient within a visualization.}
Moreover, previous studies have typically asked specific questions about simple visualizations employed for visual causal inference tasks, rather than probing users' high-level interpretation and comprehension.
In contrast, our work specifically focuses on the role of causal priors in shaping visualization interpretation.
By using variable pairs with different causal priors, established by previous work~\cite{wang2024causal}, we are able to systematically examine how both strong and weak causal beliefs affect users' overall interpretations of data visualizations.
In doing so, our results connect the literature on causal inference from visualized data to more general questions regarding visualization interpretation.

\section{Methodology}
\label{sec-method}

\begin{figure}[htbp]
    \centering
    \includegraphics[width=0.75\linewidth]{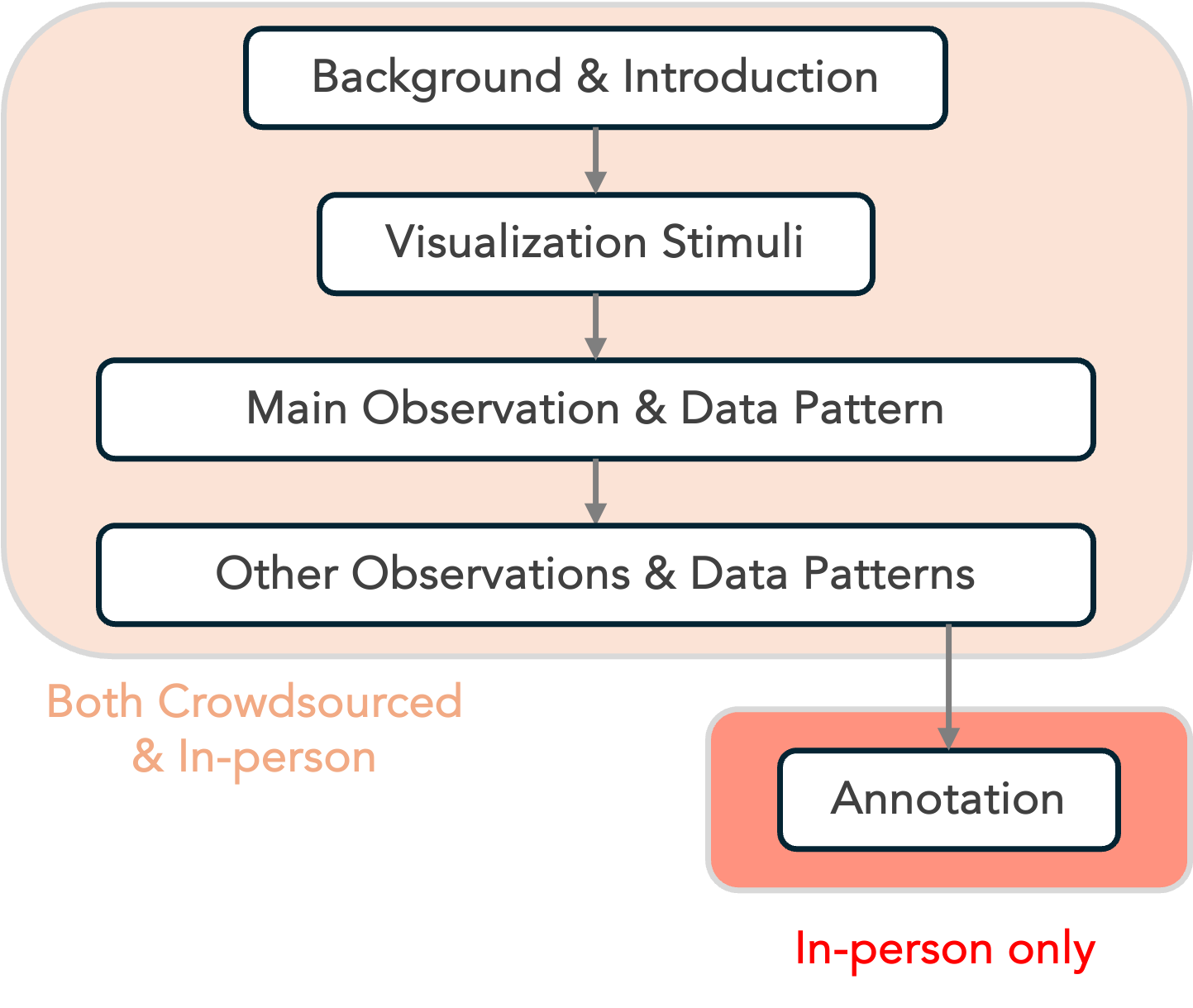}
    \vspace{-1em}
    \caption{An illustration of the overall experimental design which combined one crowd-sourced study and one in-person study to gather deeper insights into user behaviors.}
    \label{fig:exp}
    \vspace{-1em}
\end{figure}

To achieve our goal of understanding how causal priors impact visualization interpretation, we used a mixed-methods approach, \new{consisting of a pilot study to calibrate stimuli used for two distinct and complementary experiments.}
\add{First, a crowd-sourced online experiment with 60 participants was used to establish the general effects of causal priors on visualization interpretation, providing a broad base of evidence. Second, to investigate the deeper cognitive mechanisms behind these effects, we conducted an in-person, think-aloud study with 10 participants who were familiar with statistical graphs. This second study was designed to capture the rich reasoning strategies that led to participants' interpretations.}
\autoref{fig:exp} shows an overview of the experimental design.
Both studies were approved by UNC Institutional Review Board.

\subsection{Stimuli Design}

Scatterplots have been noted as the ``fruit fly'' of data visualization~\cite{rensink2013prospects} due to their common usage in a variety of studies~\cite{quadri2021survey, sarikaya2017scatterplots, szafir2023visualization}.
They have been widely employed to evaluate a diverse range of visualization-related factors, and led to a huge amount of design knowledge for visualization research~\cite{szafir2023visualization, rensink2013prospects, munzner2014visualization}, such as presenting correlation~\cite{harrison2014ranking} and clusters~\cite{wilkinson2005graph}, and analyzing high-dimensional data~\cite{van2008visualizing}.
In keeping with this previous work, for this work, we chose scatterplots for our stimuli due to their ubiquity and ability to convey multiple patterns simultaneously.

\subsubsection{Scatterplots with Ambiguous Visual Data Patterns}

We specifically focused on the two most widely used visual data patterns in scatterplots---trends and clusters~\cite{wilkinson2005graph, sarikaya2017scatterplots}---in our stimuli.
When generating stimuli, we first synthesized a set of scatterplots with different combinations of trend strength, controlled by Spearman correlation, and cluster strength, using the Gaussian norm.
In addition to these synthetic datasets, we also collected a set of real-world scatterplots from well-known datasets, such as the \emph{Iris Species}~\cite{unwin2021iris} and \emph{Palmer Penguins}~\cite{horst2020palmerpenguins} datasets.
We visually balanced the generated and selected datasets with varying numbers of points, from very sparse (around 10) to moderate (around 100) and high  (around hundreds) densities.
To explore the impact of a third variable, which is common in real-world data visualization, we also generated some scatterplots with color used to convey an additional categorical variable. Our resulting corpus included scatterplots with a single category (no color encoding), two categories, and three categories.
We first generated and collected 30 candidate scatterplots (15 synthetic, 15 from real data) and removed all data labels and variable names.

When generating and selecting the scatterplots, we attempted to achieve a similar level of strength for both the clusters and the trend, i.e., both could be weak, or both strong. The ambiguity of their visual saliency was intended to enable a more clear impact of causal priors and avoid having a single more dominant pattern, which would result in the chart being likely to be interpreted by all users in the same way~\cite{quadri2024do}.
However, a subjective assessment of the strength of patterns in a chart may introduce self-selection bias, and a robust measure that can mathematically evaluate the saliency of different visual data patterns in one chart has thus far not been developed. Existing work to this end mainly relies on eye tracking data and results in whole chart saliency maps~\cite{matzen2017data} or assesses ambiguity for one specific pattern~\cite{jeon2023clams}.

To better validate the ambiguity assessments needed for our \new{experiment}, we conducted a preliminary forced-choice \new{pilot study to calibrate stimuli} with ten users.
We showed each user the 30 candidate scatterplots and asked them to choose one of two visual data patterns as the most salient pattern in the chart: (1) trend (if a trend could be clearly observed) or (2) cluster (if two or more different classes or clusters could be clearly observed).
Based on the results, we chose nine of the 30 candidate scatterplots (three for single category, three for two categories, and three for three categories) as our final stimuli. Each of these nine charts received an equal number of votes for trend and cluster as the most salient (five for each pattern).
\new{This calibration pilot study was designed only to generate  stimuli for the subsequent experiments, and was not designed to answer any research questions itself.}
See the \href{https://osf.io/yj2wa/?view_only=9ebb800c9c5e4040a0627a6d1dade38e}{OSF supplements\footnote{\new{https://osf.io/yj2wa/?view\_only=9ebb800c9c5e4040a0627a6d1dade38e}}} for a complete collection of the final stimuli.

\add{While the \new{calibration pilot study} described above was used as a practical way to ensure a baseline level of ambiguity, we acknowledge limitations with this approach and that ambiguity may be quantified more rigorously.
Future work could \new{develop novel} computational models that measure ambiguity for specific patterns.}

\subsubsection{Choices of Data Variables}

The data variables employed in this work were selected from a previous corpus with causal priors for data variable pairs~\cite{wang2024causal}.
This data corpus was collected from a crowd-sourced population by asking directional causality judgment questions between a factor and an outcome, such as ``How much will an increase in \textit{computer speed} cause an increase in \textit{laptop price}?'' The strength of each proposed causal relationship was rated by multiple participants on a 5-point Likert scale from 1 (None) to 5 (High), and the resulting causal prior is computed as the mean of these values.
Six distinct sets of variable names were chosen for the single-category visualizations, and three sets each for the two- and three-category visualizations. More were chosen for the single-category visualizations due to their comparative ease of comprehension~\cite{quadri2024do}, resulting in more available combinations. Each set has two variables each, three of which have weak causal priors and three of which have strong causal priors.
For the multi-category scatterplots, each set has three variables each, two each of strong/weak, strong/strong, and weak/weak combinations of causal priors between the X-Y and Color-Y pairs.
\add{The color-coded category labels were assigned to a categorical dimension in the real-world datasets, such as \emph{species} from \emph{Palmer Penguins}~\cite{horst2020palmerpenguins}.}
To better examine the effect of causal priors, we chose pairs with either very weak or very strong causal prior values.
\add{The priors were drawn from a larger corpus developed through a large-scale study~\cite{wang2024causal} and filtered for low variance, indicating widespread agreement.}
\add{Strong priors were those in the top 25\% of rated causality from \cite{wang2024causal}, where a \textbf{directional} causal link is expected (e.g., \textit{study time} and \textit{grades}). Weak priors were those in the lowest 25\%, corresponding to a belief in \textbf{no plausible causal relationship in the specific given direction} (e.g., \textit{home-to-school travel time} and \textit{grades}). This distinction is crucial, as a key source of conflict in our study arises when a strong visual pattern appears for a variable pair with a weak causal prior, prompting users to explain the unexpected correlation, and vice versa.}
Some of the variables were selected to represent categorical variables with two or three values, and displayed in the color legend for multi-category scatterplots, e.g., \textit{family size} and \textit{grades}.

Prior experiments have demonstrated that causal priors can be comparable across populations in both language science~\cite{ferstl2011implicit, goikoetxea2008normative} and visualization~\cite{wang2024causal}.
To further ensure the consistency of the causal priors across users, we piloted the selected variables with five users not included in the final studies to classify each pair of variables into ``\textbf{an obvious causation}'' and ``\textbf{totally not causal}''; we only used variables that were classified into one type by all pilot users; combined with their low variance in the original corpus, these priors would therefore be expected to be consistent across our study population.
See the \href{https://osf.io/yj2wa/?view_only=9ebb800c9c5e4040a0627a6d1dade38e}{OSF supplements} for the full list of data variables.

\add{Our methodology was designed to mitigate the effect of specialized domain knowledge by selecting variable pairs whose causal priors are rooted in broad cultural norms and conventional wisdom rather than niche expertise.
While deep familiarity could modulate priors in technical domains, research suggests that such widely held, conventional beliefs are relatively stable across general populations, regardless of individual expertise in a specific area~\cite{wang2024causal}.
However, we acknowledge that a potential limitation exists that we did not explicitly filter participants based on their familiarity with topics related to the included study variables.}

\subsubsection{Final Visualization Stimuli}

We combined each of the nine scatterplots (three per category number) with all sets of variables for that category, resulting in 36 ($6 \times 3 + 3 \times 3 + 3 \times  3$) stimuli in total.
For most stimuli, we put the factor variable as the X-axis label and the outcome variable as the Y-axis label, following the convention for dependent and independent variables.
However, for three stimuli, for a single variable pair, we transposed the  X- and Y-axis labels. In this case, the outcome variable \emph{blood pressure} was on the X-axis and the factor variable \emph{alcohol consumption} was on the Y-axis. This was done in order to try to measure any impact of X- and Y-label assignment.
\autoref{fig:single} and \autoref{fig:multi} show examples of four of the final single-category and multi-category scatterplot stimuli.
See the \href{https://osf.io/yj2wa/?view_only=9ebb800c9c5e4040a0627a6d1dade38e}{OSF supplements} for the full set of stimuli.

\subsection{Experiment 1: Crowd-sourced Experiment}

In \new{Experiment 1}, we examined the potential impact of causal priors on visualization interpretation using a crowd-sourced user cohort. Previous work has demonstrated that crowdsourced qualitative results can be helpful for tasks such as visualization captioning~\cite{tang2023vistext}.

\subsubsection{Task}

\add{For \new{Experiment 1}, we designed a task with a series of questions. Before the task, we provided an introductory sentence for the scatterplot, indicating the displayed variables. We also asked participants to examine the chart before answering a small set of questions.}

\add{Users were then asked to respond to two prompts:}

\begin{itemize}
    \item \add{First, \textbf{please describe the main observation you can make about the variables in this visualization.}}
    
    \add{For this, we also asked users to describe which aspect of the chart most informed their observation and how it was informative. %
    }
    \item \add{Next, \textbf{please describe any other observations you can make about the variables in this visualization.}}

    \add{As with the first prompt, participants were asked to describe the aspects of the chart that informed their additional observations. %
    }
\end{itemize}

\add{The rationale for the multi-step procedure of asking for a main observation followed by other observations was to capture a potential hierarchy in user interpretation.
The ``main observation" prompt was designed to elicit the most salient, top-of-mind insight, which is often guided by strong visual patterns or pre-existing beliefs~\cite{wang2024causal, xiong2022seeing}.
The subsequent prompt for ``other observations" encouraged participants to look beyond this initial impression and report more subtle or secondary observations, providing a more complete picture of their overall high-level comprehension~\cite{quadri2024do}.}

\subsubsection{Participants}

We recruited a total of 60 participants for \new{Experiment 1}.
Participants all had normal or corrected-to-normal vision.
Participants were required to have at least a 95\% approval rating and IP addresses from the United States and Canada.
\new{These selection criteria were designed to maintain a consistent cultural and educational background among participants, as these factors could affect causal priors.}
The crowdsourced participants of \new{Experiment 1} included 34 males and 26 females, ranging from 22–64 years of age.
The experiment took an average of 14 minutes.
Each participant saw nine stimuli, comprising all nine scatterplots and a randomized set of variable names, and the variable names were counter-balanced such that each was seen the same number of times across all participants.
The causal prior strength of the variable names was also distributed such that each participant saw roughly the same number of strong or weak priors (4 or 5 each).
Across all participants, we received 10 responses for each single-category scatterplot stimulus and 20 responses for each multi-category scatterplot stimulus.

\subsubsection{Procedure}

Participants completed an informed consent form first and then reported their demographic information, such as age and gender.
During \new{Experiment 1}, participants completed two tasks for each of the nine trials in randomized order.
Participants answered the questions via text responses in a text box.
At the end of \new{Experiment 1}, we informed the participants that none of the visualizations contained data relations reflected in the real world.

\subsection{Experiment 2: In-person Interviews}

We observed fruitful insights on the impact of causal priors and visualization interpretation from \new{Experiment 1}.
For \new{Experiment 2}, we aimed to capture deeper insights, to which an in-person study was more amenable.

\subsubsection{Task}

The \new{Experiment 2} design was similar to \new{Experiment 1}. 
The task started with the same introductory sentence asking participants to examine the chart.

We included three tasks in \new{Experiment 2}: \textbf{main observation}, \textbf{secondary observations}, and \textbf{make annotations}, to more deeply capture participants' chart interpretations.
Given the increased flexibility afforded by an in-person study, we revised the questions for the first task to include marking aspects of the chart that informed their main observation, and indicated that marking the chart could include drawing circles, lines, labels, or other graphical elements.
The wording of the second task was revised similarly.
We asked in-person participants to \emph{mark} instead of simply describing, to more accurately and easily capture which aspects of the chart they were really attending to.
During \new{Experiment 2}, we provided a printed version of the charts, along with colored pens.

For the third annotation task, we asked participants to imagine that they were to share the chart with other people, and to add annotations (such as symbols or text) to make it easier for others to understand their observations.
The annotations were made on the printed versions of the charts using the provided pens.

In addition, during \new{Experiment 2}, we asked participants to follow a think-aloud protocol and probe to describe any points they thought were important that led to their interpretation of the chart.

\subsubsection{Participants}

We recruited a total of 10 participants for \new{Experiment 2}.
Participants all had normal or corrected-to-normal vision.
Participants were recruited via flyers, mail lists, and professional connections.
Eight were students with a university or higher degree, and two were working professionals.
All of the participants were familiar with statistical graphs and had participated in visualization, HCI, or statistics courses, or had working experience with data analysis.
The participants included 4 males and 6 females, ranging from 20–36 years of age.
We received three or four responses for each scatterplot stimulus.
\new{Experiment 2} took from 30 to 45 minutes.

\subsubsection{Procedure}

Each participant was interviewed by one or two study coordinators during \new{Experiment 2}.
Participants were first provided with the consent information and study background.
Each participant then performed the three tasks for each of the nine trials in randomized order.
Participants answered the questions following the think-aloud protocol, and the coordinator encouraged them to talk about anything they thought was important to their interpretation.
At the end of \new{Experiment 2}, we informed the participants that none of the visualizations contained data relations reflected in the real world.

\subsection{Data Coding}
\label{sec:analysis}

We collected the text responses from the online experiment, and recorded and transcribed the think-aloud interviews from in-person sessions.
\new{To derive high-level insights from the qualitative data, we conducted a thematic analysis.}
\new{Two of the co-authors independently read the participant responses from Experiment 1 and the think-aloud interview transcripts from Experiment 2 to independently capture the themes and patterns from the experiments' results.
They performed open coding on the responses, generating initial descriptive codes (e.g., ``uses causal verb leads to'', ``mentions cluster'', ``annotates question mark'', ``annotates missing variable'').
They both performed open coding on a subset of 45 overlapping responses from which the inter-rater reliability was high ($\kappa = 0.91$).
The researchers then divided and coded the remaining data using the consolidated codebook (available in the \href{https://osf.io/yj2wa/?view_only=9ebb800c9c5e4040a0627a6d1dade38e}{OSF supplements}).
During this work, the coders met regularly to collate the codes and their associated data extracts, and discuss and group codes into broader patterns.
Through this process, we identified three overarching themes that are directly relevant to our research question about the influence of causal priors on data visualization interpretation (see \autoref{tab:theme}).}

More specifically, we coded the responses for their \textcolor[RGB]{137, 172, 70}{\emph{primary observation}}, \textcolor[RGB]{137, 172, 70}{\emph{secondary observations}}, \textcolor[RGB]{255, 137, 137}{\emph{primary visual data pattern}}, \textcolor[RGB]{255, 137, 137}{\emph{secondary visual data patterns}}, \textcolor[RGB]{45, 170, 158}{\emph{annotations}}, and \textcolor[RGB]{45, 170, 158}{\emph{reasoning strategy}}.

\textbf{\textcolor[RGB]{137, 172, 70}{Primary and secondary observations}} were encoded by the level of causal attribution implied by  the causal language~\cite{hill2024casual} used by participants. Each observation was classified as \emph{explicitly causal} (e.g., \emph{cause} and \emph{lead to}), \emph{explicitly correlative} (e.g., \emph{increased/decreased} and \emph{higher/lower}), and \emph{uncertain}, such as a mediator of causal or correlate (e.g., \emph{possible that} and \emph{skeptics argue}).

\textbf{\textcolor[RGB]{255, 137, 137}{Primary and secondary visual data patterns}} were encoded into typical visual data patterns and properties found in scatterplots~\cite{wilkinson2005graph}. The two most common identified patterns were \emph{trend} and \emph{cluster}, although several others were also mentioned, including \emph{outlier}, \emph{density}, \emph{data distribution}, \emph{color}, and \emph{range}.

\textbf{\textcolor[RGB]{45, 170, 158}{Annotations}} represents the annotation results from the interview.
We first analyzed the annotated patterns using the coding from the \textbf{\textcolor[RGB]{255, 137, 137}{visual data patterns}}.
Then we categorized them into \emph{mark-based demonstration}, in which participants preferred to draw marks as annotations, \emph{simple-word illustration}, in which participants used simple words and phrases in their annotations, and \emph{detail-level explanation}, in which participants used complete sentences to explain their chart interpretations.

\textbf{\textcolor[RGB]{45, 170, 158}{Reasoning strategy}} was primarily captured from the interview results, whereby identified the reasons that participants provided to justify their observations.
More specifically, the main reasoning behaviors of the participants were categorized using the terms \emph{contextualization} and \emph{rationalization}.
\add{The codes were not mutually exclusive; a single participant response could be coded as both contextualization and rationalization if it exhibited elements of both behaviors. These instances were specifically tagged as `mixed behavior' and are discussed in \autoref{sec-behavior-mix}. Please see \autoref{sec-behavior-def} for a more detailed explanation.}

\section{Thematic Analysis}
\label{sec-theme}

The primary goal of this study was to understand how causal priors influence visualization interpretation.
Toward this aim, we applied thematic analysis to participant responses regarding their \textcolor[RGB]{137, 172, 70}{\emph{primary observation}}, \textcolor[RGB]{137, 172, 70}{\emph{secondary observations}}, \textcolor[RGB]{255, 137, 137}{\emph{primary visual data pattern}}, \textcolor[RGB]{255, 137, 137}{\emph{secondary visual data patterns}}, and \textcolor[RGB]{45, 170, 158}{\emph{annotations}}.
\add{\autoref{tab:theme} provides a high-level summary of the three main themes that emerged from this analysis:}
(1) the impact of strong causal priors on visualization interpretation, (2) the impact of weak causal priors on visualization interpretation, and (3) the similarities and differences in visualization annotation with respect to causal prior strength.
We additionally report one observation regarding whether changing the axis assignment for a variable pair has an effect on the interpretation of causality. This observation is also derived from our thematic analysis, but as it is based on data from fewer trials, we do not report the observation as an overarching theme.
This section provides an overview of our findings with key data points to support the themes. \href{https://osf.io/yj2wa/?view_only=9ebb800c9c5e4040a0627a6d1dade38e}{OSF supplements} provide more extensive quotes and examples for strong/weak causal priors and visual data patterns.
Participant quotes from the online experiment are denoted by a random ID as [\textbf{$P_{o1}$}-\textbf{$P_{o60}$}], and from the in-person experiment as [\textbf{$P_{i1}$}-\textbf{$P_{i12}$}].

\begin{table*}[htbp]
\caption{\new{Three main themes derived from the analysis, with their constituent components, indicators, and representative quotes.}}
\label{tab:theme}
\begin{tabular}{|p{2cm}|p{4.5cm}|p{5cm}|p{5cm}|}
\hline
\textbf{Main Theme} & \textbf{Components} & \textbf{Indicators} & \textbf{Representative Quotes} \\
\hline
\multicolumn{4}{|l|}{\textbf{1. Strong causal priors are associated with increased perceived causality and a focus on trend patterns}} \\
\hline
& Strong priors elicit causal language & Causal verbs (``causes", ``leads to"), low hypothetical words & ``\emph{Higher engine horsepower causes a high car price.}'' \\
\cline{2-4}
& Focused trend observation & Trend usually named as primary observation & ``\emph{more study time, higher grades.}'' \\
\hline
\multicolumn{4}{|l|}{\textbf{2. Weak causal priors are associated with lower perceived causality and more diverse pattern observations}} \\
\hline
& Weak priors suppress causal language & Uncertain language (``may have", ``maybe") and statements of no clear relationship & ``\emph{a very scattered scatterplot that doesn't show much of a relationship.}'' \\
\cline{2-4}
& Diversified pattern observations & Reports of outliers, range, density, etc. as primary observations & ``\emph{The dots are scattered... but also these few points up here are odd.}'' \\
\hline
\multicolumn{4}{|l|}{\textbf{3. Annotations reflect the strength of causal priors}} \\
\hline
& Definitive annotations for strong priors & Arrows, conclusive text, mark-based demonstrations & ``\emph{Higher smoking time leads to higher liver disorder severity.}''\\
\cline{2-4}
& Uncertain annotations for weak priors & Question marks, questioning text, labels suggesting distrust & ``\emph{Study time is (a) confounding variable?}'' \\
\hline
\end{tabular}
\end{table*}

\subsection{Theme 1: \textcolor[RGB]{148, 80, 52}{Strong Causal Priors Are Associated with Increased Perceived Causality and a Focus on Trend Patterns}}
\label{sec:t1}

\begin{figure*}[t]
    \centering
    \includegraphics[width=0.8\linewidth]{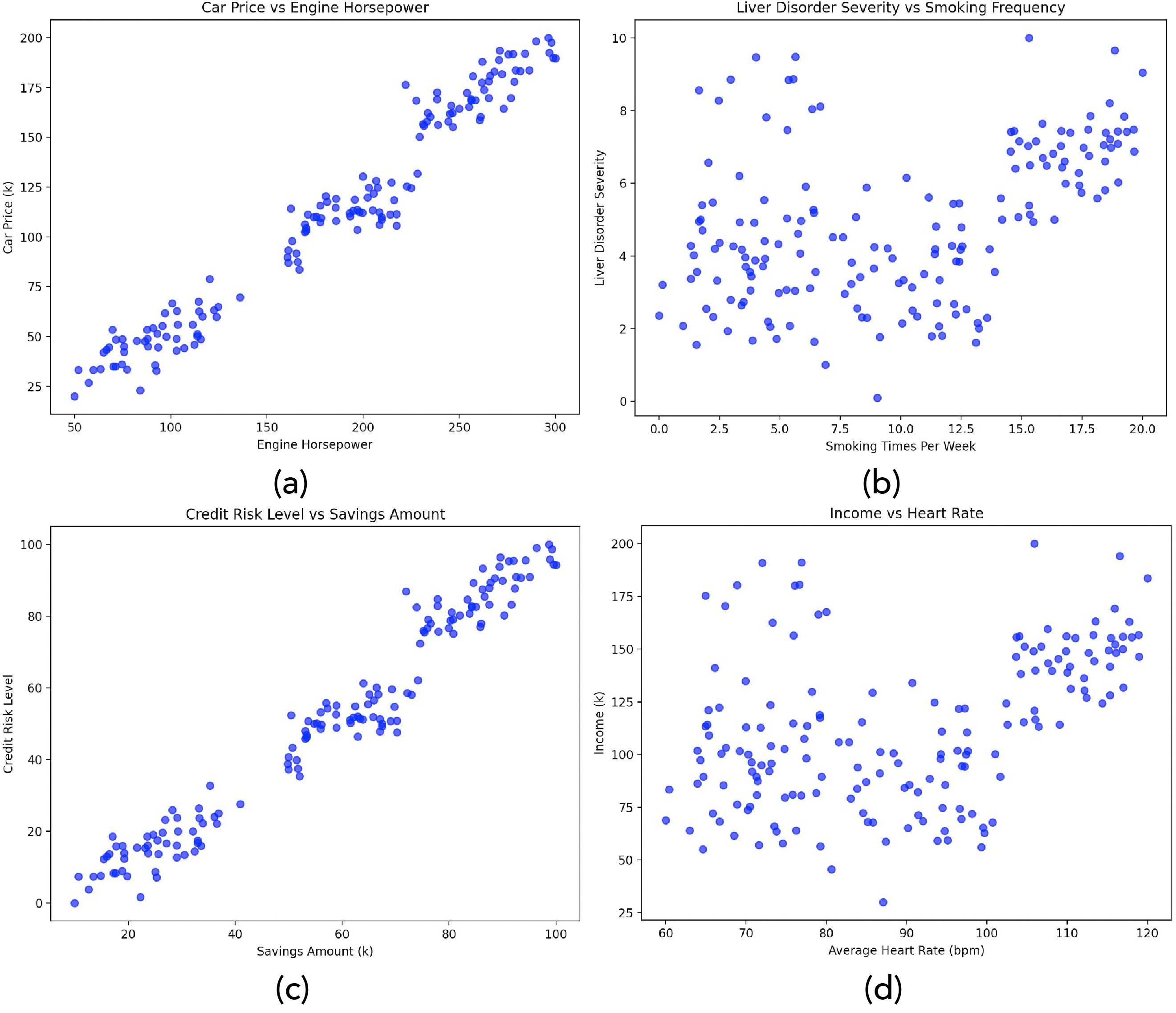}
    \vspace{-1em}
    \caption{\add{Four single-category stimuli showing different visual data patterns and variable pairs.
    Please see supplemental materials for larger images.}}
    \label{fig:single}
    \vspace{-1em}
\end{figure*}

\add{In conditions with strong causal priors, we observed that participants' reported main observations were more likely to contain consistent aspects across participants.}
This impact manifested in two different ways.
\add{First, participants tended to \textbf{interpret ambiguous visual data patterns shown in visualizations through a more causal lens}.} More specifically, 
we found that overall, 34\% of all responses (123/360) included a description of explicit causal language for visualizations that included at least one strong causal prior, \add{and an additional 9\% of them (33/360) reported similarly causal observations using less explicit, more uncertain language.}

\add{Second, participants usually \textbf{prioritized the trend pattern as their main observation when a strong causal prior was present}.}
Among these results, 89\% of the main observations (320/360) were focused on trends, and the remainder primarily on clusters (9\%, 31/360).

For single-category data visualizations with strong visual data patterns in both trends and clusters, we observed that participants are likely to describe their main observation as a trend using language indicative of a causal relationship, and are very unlikely to use uncertain language.
In their primary observations in these cases, we observed that 40\% of responses (12/30) included explicitly causal language, whereas none of the responses were reported with uncertain language.
For example, in \autoref{fig:single} (a), this single-category scatterplot shows data of \emph{car price} and \emph{engine horsepower} with obvious trend and cluster patterns. Here, more participants noted main observations such as ``\emph{Higher engine horsepower leads to higher car price}'' [$P_{o21}$] or ``\emph{I noticed that this graph shows a higher engine horsepower causes a high car price}'' [$P_{i05}$].
In the other responses, even without explicit causal language, we also observed very strong and deterministic perspectives, for example, most responses clearly stated descriptions such as ``\emph{The higher the engine horsepower, the higher the car price,}''  [$P_{o07}$], and ``\emph{What I noticed from this chart is that cars with more powerful engines usually cost more,}'' [$P_{o20}$].
All of these main observations are informed by the trend pattern.
We anticipate that in such cases, the participants tend to anchor their interpretations of visualizations in familiar causal narratives, therefore, they also elevate trend patterns as the primary evidence to support their views. Although trends directly represent correlations, previous research shows that users often infer causal relationships from visualizations of trends~\cite{xiong2019illusion, wang2024causal}.
In their secondary observations, we observed a focus on the clustering patterns, but with a much larger range of patterns reported.
For example, we observed ``\emph{There is a decent gap in cars that have engine horsepower of 140 and 160,}'' which is informed by ``\emph{...data points are spread out on the chart, especially any groups or spaces...}'' [$P_{o13}$].
These observations align with how we designed these charts by introducing trend and cluster patterns with equal visual strength.
However, we also found that in some cases users described other aspects of the data, such as the \emph{range} of data values, ``\emph{No car less than 100 thousand dollars has more than a 150 horsepower,}'' informed by ``\emph{There is no dot at or below the 100k line that is past the 150 engine horsepower line,}'' [$P_{o36}$].

For single-category visualizations with strong causal priors but weak visual data patterns, we also observed a similar tendency to report using causal language, however, some uncertain language was also observed.
27\% of responses (8/30) used causal language and 7\% of responses (2/30) were stated with uncertainty.
For example, \autoref{fig:single} (b) shows a visualization of \emph{liver disorder severity} and \emph{smoking frequency} with weak patterns for both trend and cluster.
Even though the patterns are weak, many participants still noted their main observations using causal language. However the language was often less certain, such as ``\emph{liver disorder severity may be affected by smoking frequency,}'' [$P_{o44}$], ``\emph{Seems like more smoking times per week meant a higher chance of liver disorder,}'' [$P_{o28}$], and ``\emph{More smoking times a week can cause more severe liver disorder,}'' [$P_{i01}$].
These results show that, with strong priors, even with relatively weak visual patterns, participants still interpret the visualized data as indicative of a causal relationship.
In their secondary observations, however, we did not observe a main focus on clusters, but rather a large variety of other observations.
Their secondary observations include those related to density---``\emph{The dots are scattered a lot especially with the smoking times per week it seems kind of all over the place,}'' [$P_{o09}$], range---``\emph{if you smoke more than 15 times per week and if you have a liver disorder than it is likely to be worse than level 4 severity,}'' [$P_{o39}$], and outliers---``\emph{Some people who smoke less have high liver disorder severity,}'' [$P_{o01}$]. Weak visual data patterns without causal priors to influence the interpretation seem more likely to elicit a diverse range of perspectives and focuses when interpreting them~\cite{bearfield2024same, quadri2024do}.

\begin{figure*}[t]
    \centering
    \includegraphics[width=0.8\linewidth]{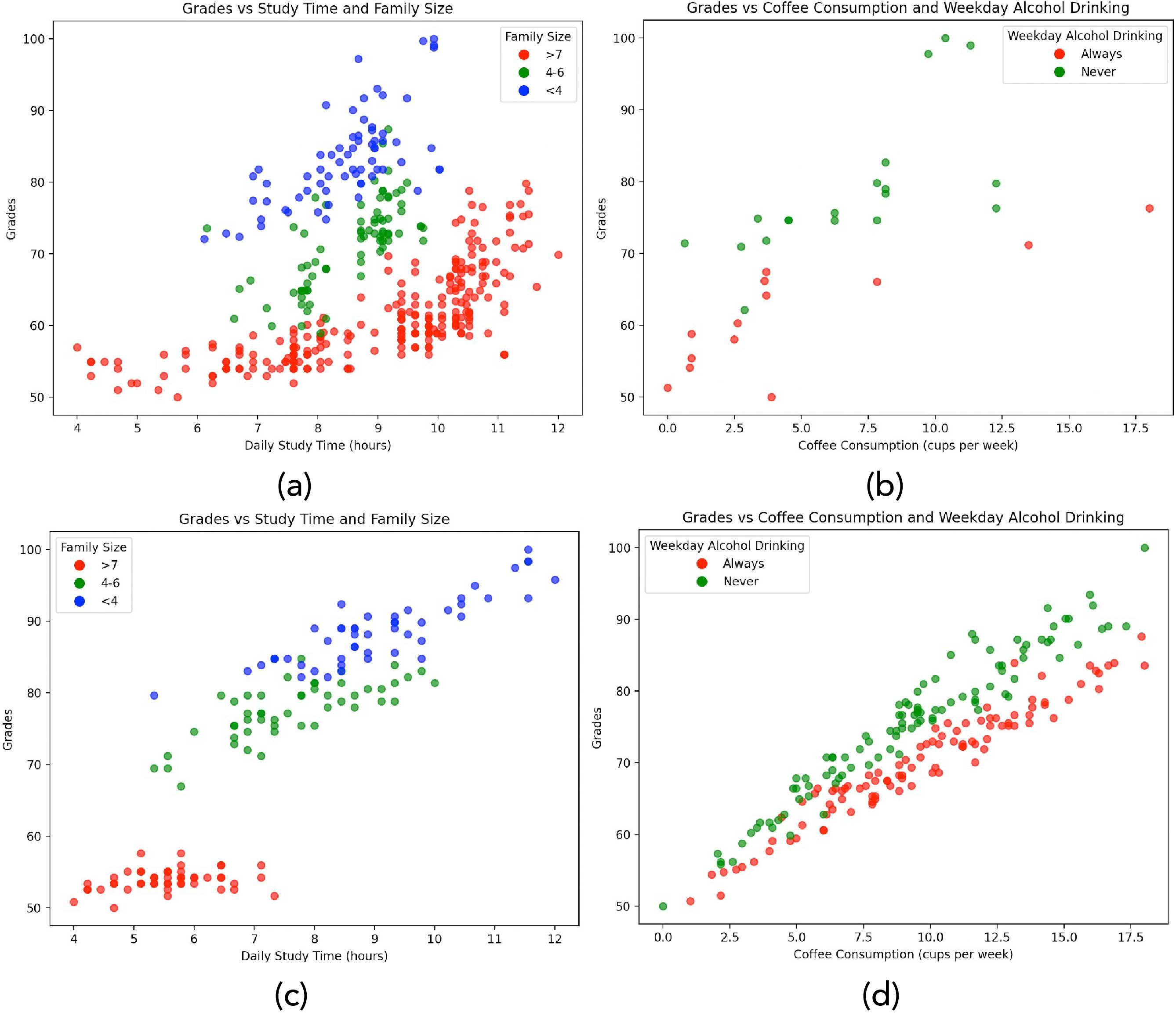}
    \vspace{-1em}
    \caption{\add{Four multi-category stimuli showing different visual data patterns and variable pairs.
    Please see supplemental materials for larger images.}}
    \label{fig:multi}
    \vspace{-1em}
\end{figure*}

In the cases above, we found that users prioritized trends over clusters in single-category visualizations.
However, the interpretation of multi-category visualizations may be more complicated.
For example, we found participants complained that the increased complexity of a multi-category visualization would make it harder to interpret, for example, ``\emph{...it basically measures many different things at once...This ends up making things a little bit unpredictable,}'' [$P_{o50}$].
However, we still saw a primary effect similar to that of single-category visualizations, that if the data variables shown on the X- and Y-axis have a strong causal prior, users' main observations also focused on the trend pattern.
For example, \autoref{fig:multi} (a) shows \emph{grades} (Y-axis), \emph{study time} (X-axis), and \emph{family size} (color legend).
The causal prior for \emph{study time} to \emph{grades} is strong, whereas the prior for \emph{family size} to \emph{grades} is weak.  
Therefore, almost all of the main observations (18/20) focused on the trend related to study time, such as, ``\emph{Higher daily study time is correlated with much higher grades,}'' because ``\emph{The higher Y axis values means study more cause higher grades, maybe a better understanding of concepts,}'' [$P_{o29}$].
In this case, despite the additional variable, a strong causal prior prioritized the trend observation.

On the contrary, we found that when a strong causal prior is present for the color legend variable in a visualization,
a relatively large number of main observations indicate the clustering pattern.
For example, \autoref{fig:multi} (b) shows \emph{grades} (Y-axis), \emph{coffee consumption} (X-axis), and \emph{weekday alcohol drinking} (color legend).
In this case the strong causal prior for \emph{weekday alcohol drinking} on \emph{grades} resulted in 60\% of the main observations (12/20) being cluster. In this case we observed more responses focused on the differences between categories.
For example, ``\emph{People who drink alcohol on weekdays have worse grades,}'' [$P_{o08}$] and ``\emph{Drinking alcohol during week days hurts grades,}'' [$P_{o11}$].
The strong causal prior was often associated with shifting the main focus from the trend to the differences between categories.
However, many users (7/20) still focused on the overall X-Y trend.
This may be partly because people often feel that the relation between the X and Y axes may be the most important aspect to focus on in a chart~\cite{szafir2023visualization, munzner2014visualization}.

\subsection{Theme 2: \textcolor[RGB]{79, 28, 81}{Weak Causal Priors Are Associated with Lower Perceived Causality and More Diverse Pattern Observations}}
\label{sec:t2}

\add{In conditions with weak causal priors, we observed that participants' interpretations were typically less consistent across users compared to those for strong causal priors.}
\add{In such cases, we found that users were more likely to \textbf{describe their interpretations in a non-causal way}, and reported \textbf{a more diverse range of observed visual data patterns}.}
We identified less than 3\% of the responses (5/180) that included causal language, whereas 42\% (76/180) included uncertain language.
We found a diverse set of main observations, including trend (31\%, 56/180), cluster (25\%, 45/180), outliers (22\%, 40/180), range (18\%, 32/180), and density (3\%, 5/180).

We did not find any explicit causal language in responses for weak causal priors.
For single-category visualizations with strong visual data patterns, even though users still focused on trends, causal language was never used in their observations.
For example, \autoref{fig:single} (c) describes exactly the same visual data patterns as (a), but using a different variable pair with a weak causal prior, \emph{credit risk level} and \emph{savings amount}.
In the main observations of this chart, we did not observe any obvious causal language from users' responses, however, they did focus on the trend pattern and used less certain language, such as ``\emph{maybe a relation between more savings and high credit risk,}'' [$P_{o10}$].
We also observed cases where users tried to reason beyond the data in this chart when describing why this pattern was informative for their observations, for example, ``\emph{... some high risk people may have a luxury life or live in expensive cities...}'' [$P_{o01}$]. This type of behavior was not observed for responses in \autoref{fig:single} (a).
In such cases, users may have struggled to interpret the chart directly because the weak causal priors did not align with the strong visualized patterns. Therefore, most online users simply described the pattern, while others tried to imagine reasons beyond the chart to make it seem more rational.

For weak causal priors with weak visual stimuli, unlike cases with strong causal priors, we mostly observed clear and certain language regarding observations of an uncertain trend or non-correlated pattern.
\autoref{fig:single} (d) visualizes the same visual data patterns as (b) but replaced with variable names with weaker causal priors,  \emph{heart rate} and \emph{income}.
In this case, all participants (10/10) were certain that there was no correlation in their main observation. For example, ``\emph{There isn't that much of a correlation,}'' [$P_{o09}$] and ``\emph{a very scattered scatterplot that doesn't show much of a relationship between heart rate and income}'' because ``\emph{The dots are randomly and widely distributed,}'' [$P_{o22}$].
These observations differ from those for \autoref{fig:single} (b).
A potential explanation is that users did not have a strong causal prior for the data variables, which therefore did not strengthen the subtle trend in their mind.
We also observed that a larger number of different types of patterns were reported as main observations by participants, for example, \emph{range} was reported three times for \autoref{fig:single} (c)---``\emph{The savings amount spanned a large range,}'' [$P_{o01}$], and \emph{outliers} was reported four times for \autoref{fig:single} (d)---``\emph{There are few people who make under 40k,}'' [$P_{o38}$].
Such observations typically appeared in the secondary observations in the strong causal prior cases.

For multi-category visualizations, we found that whenever a weak causal prior was present, its pattern was less likely to appear in the main observation.
For example, \emph{family size} appeared in 10\% (2/20) of the main observations in \autoref{fig:multi} (a) and \emph{coffee consumption}
appeared in 25\% (5/20) of the main observations in \autoref{fig:multi} (b).
Further, the observations of weak causal priors were influenced by particular differences in visual data patterns.
\autoref{fig:multi} (c) describes the same variable pairs as (a), but has stronger visual data patterns, especially the separation between red and the other two categories.
In this case, we found 40\% (8/20) of the main observations included \emph{family size}, for example, ``\emph{Family size of 4 or less usually have better grades than others,}'' [$P_{o13}$], and ``\emph{The lower the family size the higher the grades can be,}'' [$P_{o21}$].
Similarly, \autoref{fig:multi} (d) describes the same variable pairs as (b), but has stronger trend patterns.
In alignment with this, 60\% of participants' responses (12/20) centered around the relation between coffee drinking and grades, such as ``\emph{grades are positively correlated with coffee consumption}'' informed by ``\emph{The green and red dots going in an upward trajectory the more coffee consumption per week shows me this is true,}'' [$P_{o53}$]. 
An additional factor could be the increased complexity of multi-category visualizations, which may result in an increased diversity of interpretations. This aligns with what we found in user responses as well as findings from previous work~\cite{quadri2024do}.

\subsection{Theme 3: \textcolor[RGB]{95, 139, 76}{Annotations Reflect the Strength of Causal Priors}}
\label{sec:t3}

\begin{figure*}[t]
    \centering
    \includegraphics[width=\linewidth]{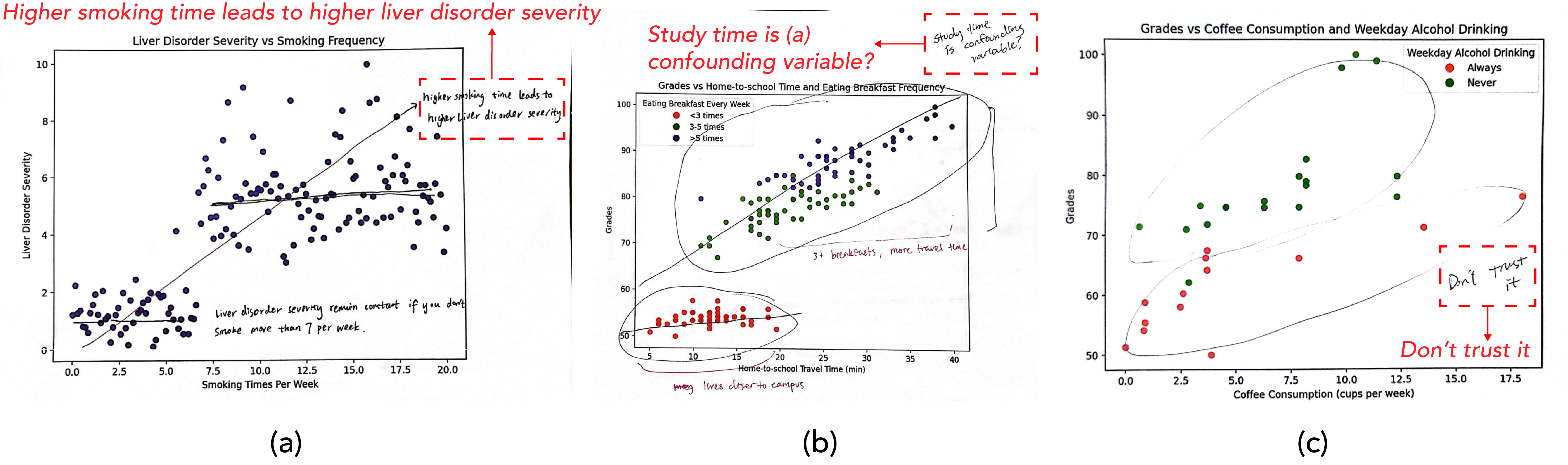}
    \vspace{-1em}
    \caption{Images of annotation results from three users.
    Highlighted annotations are as follows: (a) ``\emph{Higher smoking time leads to higher liver disorder severity}''; (b) ``\emph{Study time is (a) confounding variable?}''; (c) ``\emph{Don't trust it}''.}
    \label{fig:annotation}
    \vspace{-1em}
\end{figure*}

We also analyzed user annotations.
First, we found that the annotations were all \textbf{aligned with their main observations} but \textbf{reflected the strength of the causal priors}.
Specifically, when there were variable pairs with strong causal priors,  annotations were more likely to convey more certainty about the evidence for the visual data pattern, for example, in \autoref{fig:annotation} (a), participant [$P_{i01}$] annotated a weak trend pattern with a trend line and labeled it as ``\emph{Higher smoking times lead to higher liver disorder severity}'' (indicated by the red box).
However, when the variable pairs had weak causal priors, participants conveyed less certainty, even with strong patterns. For example, in \autoref{fig:annotation} (b), the relations between \emph{home to school travel time}, \emph{eating breakfast frequency}, and \emph{grades} was questioned by annotator [$P_{i09}$] and labeled as ``\emph{Study time is (a) confounding variable?}'' (indicated by the red box).
We even found that some extremely negative attitudes were expressed via annotations, for example, participants [$P_{i07}$] added the label ``\emph{Don't trust it}'' on the visualized positive trend between \emph{coffee consumption} and \emph{grades}, shown in \autoref{fig:annotation} (c).
We also saw noticeable individual differences with respect to annotation style; however, annotation styles are not the core focus of this paper, please see the supplemental materials for relevant discussions.

\subsection{Observation: \textcolor[RGB]{210, 159, 128}{People May Not Distinguish the Direction of Causality based on the Assignment of X and Y-Axes}}

Another noteworthy finding from our study is that participants did not reliably use axis assignments to infer the direction of causality. Causality is a directional relation between concepts~\cite{pearl2009causality}, and causal priors can reflect this directionality. For example, the perception of more \emph{study time} causing higher \emph{grades} does not mean that the higher \emph{grades} are viewed as causing more \emph{study time}. Previous studies of causality perception in visualization have typically followed a consistent approach in which the X-axis variable is seen as causing (or not) changes in the Y-axis variable~\cite{xiong2019illusion, wang2024causal}.  In other words, the typical assumption is that X causes differences in Y. This aligns with the common perception that the X-axis represents an independent variable~\cite{szafir2023visualization, munzner2014visualization}. 

We adopted a similar convention for most of the stimuli in our studies. However, to investigate the impact of axis assignment on the directionality of perceived causality, we also included three stimuli that followed a transposed design. All three transposed stimuli displayed the same concept pair (\emph{alcohol consumption} and \emph{blood pressure}) with a strong unidirectional causal prior, but with different variations of visualized data. In all cases, the causal prior's independent variable \emph{alcohol consumption} was assigned to the Y axis while the causal prior's dependent variable \emph{blood pressure} was assigned to the X-axis. Figure 1 in the supplemental material includes these stimuli.
\add{Because this variation adds a layer of complexity to the study design, its inclusion was intentionally limited to a small set of stimuli to test for this potential interaction without disrupting the primary focus of the study.}

We found that for these transposed stimuli, all participants explained alcohol consumption as an independent variable despite it being assigned to the Y-axis. This suggests that the users prioritized alignment with their causal priors more than the axis assignment. For example, ``\emph{I noticed that drinking more alcohol seems to be linked to higher blood pressure,}'' [$P_{o43}$], and ``\emph{The more you drink the higher your blood pressure will be,}''  [$P_{o07}$]. 
These results suggest that the causal priors are influential and that the topic warrants further investigation.
\add{However, we also acknowledge that this is based on a smaller number of observations, with the manipulation of the X-Y axis assignment included as a limited, exploratory probe rather than as a core component of the design.}

\section{Visualization Interpretation Behaviors}
\label{sec-results}

To understand more fully how causal priors shape visualization interpretation, we investigated users' \textcolor[RGB]{45, 170, 158}{\emph{reasoning strategies}}.
\add{The findings in this section are drawn primarily from the rich, qualitative data gathered during our in-person think-aloud study (N=10).}
Based on our observations, described in more detail in \autoref{sec-behavior-def}, we categorized users' reasoning strategies into two archetypal behaviors: \emph{contextualization} and \emph{rationalization}.
\autoref{fig:teaser} shows how users' visualization interpretation behaviors vary by the strengths of causal priors and visual data patterns.
See the \href{https://osf.io/yj2wa/?view_only=9ebb800c9c5e4040a0627a6d1dade38e}{OSF supplements} for additional quotes and examples that provide further evidence for each of these behaviors.

\subsection{Two Archetypal Behaviors}
\label{sec-behavior-def}

The cognitive processes involved in visualization interpretation are complex and relevant to psychophysical behaviors~\cite{wang2025characterizing}. However, by examining the results of our studies, we noticed two general overarching behaviors when users were interpreting our visualization stimuli.
These behaviors exist on a spectrum, and if viewed in isolation, are an oversimplification of the complex underlying cognitive processes. However they provide a useful lens for examining certain aspects of visualization interpretation in the context of causal priors. 

\new{To understand how causal priors shape interpretation, we focused on the reasoning strategies participants used to justify their observations
Through an inductive process, we analyzed the open-coded justifications and found two dominant but contrasting patterns that emerged from how participants reconciled (or failed to reconcile) the visualized data with their implied causal priors.}

\new{The first pattern is integrative. Many justifications involved participants accepting a salient visual pattern and merging it into a narrative informed by their causal priors. They used their priors as a lens to explain or enrich the pattern, often by adding domain-specific context not present in the chart.
We observed that this pattern was most prevalent when the strengths of the causal prior and the visual pattern were consistent (both strong or both weak).}

\new{Based on this, we identified} the first interpretation behavior, \textbf{\emph{contextualization}},
describing situations in which users \add{\textbf{accept the visual patterns}} derived from visualized data that align with their prior beliefs of causality.
\add{With contextualization, users apply their prior knowledge to add meaning and explanation to the visualized data. For example, a user who believes horsepower causes higher car prices might accept a positive trend and use their knowledge of car types (e.g., ``\emph{economy cars} vs. \emph{race cars}") to explain clusters within that trend.}

\new{The second pattern involves discounting visualized information. Participants were seen to reject or express skepticism toward salient visual patterns that conflicted with their causal priors. To resolve this dissonance, they hypothesized external reasons, such as unmeasured variables or data quality issues, to explain away the pattern.
This pattern was most prevalent when the causal prior and the visual pattern disagreed (one strong and the other weak).}

\new{Based on this, we identified} the second behavior, \textbf{\emph{rationalization}},
describes situations in which users attempt to \add{\textbf{discount or explain away observed visual data patterns}} when the visualized data has obvious differences from their prior beliefs of causality.
\add{When rationalizing, users often propose external factors not present in the chart, such as confounding variables, data selection bias, or missing data, to justify why the observed pattern is misleading or shouldn't be trusted.}

While users might exhibit different behaviors depending on the specifics of a given chart, each behavior typically appeared individually at a given point in time. However, in some cases users provided reasons that relied on both types of behavior. This reflects our earlier comment that the behaviors exist upon a spectrum.

\subsection{Contextualization: Accepting Observations in Line with Causal Priors}
\label{sec:b1}

We found that contextualization commonly occurs in cases where the strength of the causal prior and the strength of the ambiguous visual data patterns were consistent: either both strong or both weak.
We also found that contextualization typically resulted in more certain conclusions and higher trust.
Contextualization was observed for all in-person participants.
People also contextualized to explain the secondary visual data patternsin the chart beyond their major observation.

For example, in \autoref{fig:single} (a), the variables have a high causal prior and both the trend pattern and cluster pattern are strong. In this case, one online participant [$P_{o13}$] explained the observed gaps across clusters as: ``\emph{These either mean they are less frequently made or the material cost begins to significantly jump once a vehicle is over 160 engine horsepower.}''
Participant [$P_{i01}$] viewed this as: ``\emph{I agree...fits into my assumption, there are different types of cars, the highest dots may be race cars, the lowest points may be economy car...}''
In both cases, we observed users trying to contextualize the visualized data within their own prior beliefs, in this way they explained the other observed differences and visual data patterns from charts.

\subsection{Rationalization: Rejecting Observations and Thinking Outside the Box}
\label{sec:b2}

On the contrary, rationalization was more commonly described by participants when they saw visualizations with a significant disparity between the strength of causal priors and the strength of visual data patterns.
In these cases, users tended to draw more uncertain conclusions and exhibit lower trust in the observed data patterns.
Like contextualization, rationalization was also identified in all in-person participants.
We found that the most common expressions of rationalization were assuming the existence of an additional variable beyond what was shown in the chart that provided an alternative explanation for the observed visual data pattern, or that data were selected in such a way as to be responsible.

For example, \autoref{fig:single} (c) includes a visualization with strong visual data patterns but variable pairs with a weak causal prior.
Though reporting the trend pattern as the primary observation, participant [$P_{i08}$] noted the reasoning behind this pattern as ``\emph{I was shocked by this... But the increased credit risk may not be because of savings amount. I would rather imagine these people with higher savings also do more risky investments, people with low savings do have these changes so they cannot lose... However, it is not common, these data may be selected from people prefer high risk.}''
For a multi-category stimulus showing a clustering pattern for \emph{gender} and \emph{cholesterol} with weak causal priors (see Figure 2 (a) in the \href{https://osf.io/yj2wa/?view_only=9ebb800c9c5e4040a0627a6d1dade38e}{OSF Appendix}), participant [$P_{i02}$] said ``\emph{I don’t think gender make sense to cholesterol. Maybe this data is from patients with a relevant disease?}''
In another multi-category example with a strong pattern but with weak causal priors between \emph{home-to-school travel time}, \emph{eating breakfast frequency}, and \emph{grades} (see Figure 2 (b) in the \href{https://osf.io/yj2wa/?view_only=9ebb800c9c5e4040a0627a6d1dade38e}{OSF Appendix}), participant [$P_{i05}$] noted ``\emph{I think this [the increasing trend] is unreal. This could be due to a lack of data points.}'' For this same example,  participant [$P_{i09}$] said ``\emph{I don't think there's any effects...This may be due to the school district system; people from far away outside of the districts may have to work harder to get enrolled in and their parents also take more care of them.}''
Rationalization can also appear when people try to explain an unexpected pattern, such as  [$P_{i09}$] reflecting on the points on the right side with both high heart rate and high income in \autoref{fig:single} (d), ``\emph{Maybe people with high income also do more fitness which changes their heart rate, but that's not actually an effect of heart rates.}'

\subsection{Mixture of Interpretation Behaviors}
\label{sec-behavior-mix}

We also found examples where different interpretation behaviors appeared in the same reasoning process on the same chart by the same person.
We identified this mixed behavior in six in-person participants.
For \autoref{fig:multi} (c), participant [$P_{i01}$] explained the reasoning for their main observation as ``\emph{It's quite convincing...studying more will give you a better grade. If we can sample more data points [the red dots], I assume it will still present a similar increasing trend.}''
In this case, this participant contextualized the main visual data patterns but also observed that data points colored in red have a relatively smaller number and rationalized it by imagining if the chart had more points, it would introduce the same trend.
When both the causal priors and visual data patterns are weak, we also found similar interpretations, such as in \autoref{fig:single} (d), participant [$P_{i03}$] said ``\emph{The data looks good to me, I would say there's not really a trend...[points on the right top side] These are like outliers.}''
The participant contextualized the \emph{not really a trend} from the overall data, which aligned with the causal prior, however, the \emph{points on the right top side} were rationalized as outliers instead of part of the primary visual data pattern.

\subsection{I Don't Trust It: A Corner Case}
\label{sec-trust}

Furthermore, we identified a corner case in which participants neither contextualized nor rationalized, instead responding that the visualization was not trustworthy.
This corner case was only captured in situations where the causal prior was weak but the strength of visual data patterns was strong
Three in-person participants exhibited this behavior at some point.
For example, \autoref{fig:annotation} (c) shows an increasing trend between \emph{coffee consumption} and \emph{grades}. [$P_{i07}$] noted that ``\emph{I cannot trust this chart}'', both because it conflicted with their causal priors and due to a small number of data points.
Similarly, Figure 2 (c) in the \href{https://osf.io/yj2wa/?view_only=9ebb800c9c5e4040a0627a6d1dade38e}{OSF Appendix} shows a strong trend  between \emph{exercise frequency} and \emph{cholesterol}. [$P_{i02}$] noted that ``\emph{I have less than 10\% confidence and trust on this observation, I cannot believe exercise will cause higher cholesterol.}''
We expect that such corner cases may be due to a lack of trust in certain data visualizations because the visual data patterns greatly from their causal priors.

\section{Discussion}
\label{sec-discussion}

We investigated how pre-existing beliefs regarding causal relationships can affect data visualization interpretation.
Our results reveal that interpretations can vary based on the relationship between the strength of causal priors and the strength of data patterns shown in the visualization.
We anticipate that these results may provide opportunities to view previous visualization research from more causal perspective.

\subsection{Reflection on Previous Findings}

Our results relate to various insights from previous research.
Past work focused on high-level visualization comprehension noted the complexity of the process, and that results from traditional experiments may not accurately predict people's comprehension~\cite{quadri2024do}.
Our work expands upon previous insights on how prior beliefs impact the perception of causality~\cite{wang2024causal} and correlation~\cite{xiong2022seeing}, demonstrating their impact on more complex visualization interpretations.

In other relevant work, researchers found that the same data could lead to different interpretations or mental models~\cite{bearfield2024same, williams2023data}.
Our results align with these findings, showing that such divergences can appear via the integration of bottom-up perceptual cues with top-down cognitive processes that rely on pre-existing causal priors.
The consistency of these experimental results suggests that human subjects' underlying cognitive processes significantly impact how data is interpreted.

Previous work has also shown the impact of prior beliefs in \add{confirmation bias within correlation and causality judgments~\cite{xiong2022seeing, wang2024causal, li2025confirmation}, user attitudes~\cite{rogha2024impact}, and public receptivity~\cite{he2023enthusiastic}.}
Our result also reflects that such biases may play a role in how individuals process visualized data, as seen in the \textit{rationalization} behavior where people selectively interpret or explain away certain visual cues, leading to an overemphasis on data patterns that confirm their preconceptions, while potentially ignoring contradictory evidence.

\new{Our findings on how agreement, or lack thereof, between causal priors and visual patterns drives interpretation behaviors (contextualization vs. rationalization) resonate with similar frameworks in the study of correlation judgments.
For instance, Karduni et al.~\cite{karduni2020bayesian} employed a Bayesian model to show how prior beliefs about correlations are updated when presented with congruent or incongruent visual evidence from scatterplots.
Xiong et al.~\cite{xiong2022seeing} reported \emph{belief bias}, demonstrating how pre-existing beliefs of correlation cause people to see correlations that aren't there (or overestimate them) if they want to see them.
While their work focuses on the perception of association strength, our study demonstrates a similar phenomenon for causal inferences.
This suggests a broader cognitive mechanism in which prior beliefs, whether about correlation or causation, interact with visualized evidence, guiding interpretation towards either assimilation (contextualization) or dismissal (rationalization).}

\new{While our use of population-aggregated causal priors enabled us to identify a shared cognitive effect, it overlooks the role of individual differences, especially the idiosyncrasy of beliefs.
Participants likely varied not just in the direction of their priors (aligned with our strong/weak categorization) but in the certainty~\cite{duke2018preference}, polarization~\cite{li2025confirmation}, and source of that belief~\cite{smith1991belief}.
A prior held with deep personal conviction (e.g., different levels of professional expertise or experience~\cite{burns2023we}) might amplify the observed behaviors, leading to more robust contextualization or more inventive rationalization.}
For example, the archetypal interpretation behaviors of contextualization and rationalization may also reflect different levels of trust in data visualization interpretation through the lens of causality.
\new{More specifically, our observed corner case of outright distrust (\autoref{sec-trust}) may be a precursor to behaviors more common in such contexts.
Future studies should incorporate measures of belief certainty and personal relevance to model these individual factors more explicitly, investigating potential interpretation behaviors beyond two archetypes.}

\new{To analyze causal priors in a study without eliciting these individual causal priors, we relied on priors that were assumed to be relatively consistent across the study population.
In contrast, in contexts where widely shared prior beliefs are absent, such as controversial or niche subjects, it could be particularly important to identify and study individual measures of belief certainty and personal relevance.
In such cases, the influence of visual salience~\cite{quadri2024do}, narrative framing~\cite{hullman2011visualization}, or even emotions~\cite{pekrun2022emotions} may compete with causal priors as another driver of interpretation.
Investigating visualization interpretation in these high-conflict or polarized topics~\cite{li2025confirmation} could be an important future direction for understanding data communication across different settings.
}

\subsection{Limitations}

\add{Due to the exploratory nature of this study, the sample size is modest. The insights should be considered preliminary and will require further validation through follow-up studies with larger cohorts of users.}
In addition, this study focused exclusively on scatterplots, which may demand higher cognitive processing~\cite{cleveland1984graphical, quadri2024do} than other chart types.
Future work should also include a wider range of chart types and design variations.
\add{Another limitation is that the variables tested in this study were associated with directional causal priors. However, in some cases, a weak causal prior for variables in one direction might be paired with a strong causal prior in the reversed direction. 
Further work should consider this bi-directional nature of causality.}

\add{Finally, while we generated stimuli based on the strength of causal priors and visual data patterns, our approach lacks a theoretically-grounded mathematical metric to describe these factors.
Future research should explore the development and application of such metrics to better model these factors. Relatedly, our study used population-derived causal priors with low variance between people rather than unique individual causal priors obtained directly from participants. Future work could explore more sophisticated study designs that incorporate individualized participant-provided priors, though care must be taken to avoid biasing the participants' responses to the study tasks through the elicitation of their individual priors.}

\section{Conclusion}
\label{sec-conclusion}

This work explored the impact of causal priors on high-level data visualization interpretation.
We introduced a mixed design study with both crowd-sourced and in-person participants.
Our experiments viewed the visualization interpretation problem via a lens of causality.
By studying the interpretation of scatterplots with ambiguous trend and cluster visual data patterns, our results suggest that causal priors---a priori beliefs about causal relationships---affect aspects of interpretation, including observations and annotations.
More specifically, we found that strong causal priors and weak causal priors lead to different impacts on perceived causality, people's observations, and annotations.
Further, we observed two archetypal visualization interpretation behaviors, \emph{contextualization} and \emph{rationalization}, related to the strengths of both causal priors and visual data patterns.
These results connect high-level interpretation and visual causal inference, and highlight the opportunities to view visualization interpretation via the lens of causality.

\bibliographystyle{IEEEtran}
\bibliography{main}

\vspace{-1em}

\end{document}